\documentclass{elsart}

\usepackage{graphicx}

\begin{document}

\begin{frontmatter}



{\raggedright
\noindent \textbf{Preprint of:}\\
Timo A. Nieminen\\
``Comment on ``Geometric absorption of electromagnetic angular momentum'',
C. Konz, G. Benford''\\
\textit{Optics Communications} \textbf{235}(1--3) 227--229 (2004)\\
\hrulefill
}

\title{Comment on ``Geometric absorption of electromagnetic angular momentum'',
C. Konz, G. Benford}

\author{Timo A. Nieminen}

\ead{timo@physics.uq.edu.au}

\address{Centre for Biophotonics and Laser Science, Department of Physics, The
University of Queensland, Brisbane QLD 4072, Australia}

\begin{abstract}
The core theorem on which the above paper is centred---that a perfectly
conducting body of revolution absorbs no angular momentum from an
axisymmetric electromagnetic wave field---is in fact a special case of a more
general result in electromagnetic scattering theory. In addition, the scaling
of the efficiency of transfer of angular momentum to an object with
the wavelength and object size merits further discussion. Finally,
some comments are made on the choice of terminology and the
erroneous statement that a circularly polarized plane wave
does not carry angular momentum.
\end{abstract}

\end{frontmatter}

\thispagestyle{empty}

\section{Symmetry and the transfer of angular momentum}

The authors of the above paper~\cite{konz2003} state a theorem:
\begin{quote}
... a perfectly conducting body of revolution with
a piecewise smooth surface around the axis of symmetry (for instance a cone,
a disk, a cylinder, or a sphere) absorbs \emph{no} angular momentum $\vec{L}$
from an axisymmetric electromagnetic wave field.
\end{quote}
This is in fact a special case of a more general result in electromagnetic
scattering theory, namely that a scatterer that is rotationally symmetric
about the $z$ axis does not couple different azimuthal
modes~\cite{waterman1971}, that is, it does not couple modes of differing
angular momentum about the $z$ axis. This being the case, no angular
momentum about the $z$ axis will be transferred to a nonabsorbing
axisymmetric scatterer, \emph{regardless of the structure of the incident
field}.

This result is weaker than the theorem stated by the Konz and
Benford~\cite{konz2003}, since only the $z$ component of angular momentum
is considered, but is more general, applying to any nonabsorbing scatterer.
To generalize Konz's result to dielectric scatterers, it is sufficient to
consider symmetries of the incident field and the scatterer which lead to
zero torque on the scatterer about other axes. For example,
\begin{itemize}
\item mirror symmetry of the beam and scatterer about the $xy$ plane
\item mirror symmetry of the beam about any plane containing the $z$ axis
\item rotational symmetry of the beam about the $z$ axis
\item rotational point group symmetry of the beam about the $z$ axis
\end{itemize}
all result in zero torque about the origin. Clearly, rotational symmetry of
the incident field is not required. It is difficult (and beyond the scope
of this comment) to exactly and completely
state the conditions under which the total torque about the origin will be
zero. However, it should be noted that only very special types of
fields---radially and azimuthally polarized beams---are in fact axisymmetric,
if one considers the symmetry of the electric and magnetic fields. Broader
definitions of symmetry, considering the symmetry of the Poynting vector, or
the energy density, can be used, in which case, a wider class of
fields can be considered to be axisymmetric. Fields that fail to satisfy
any of these criteria can still produce zero torque---for example, higher
order Gaussian beams.

\section{Scaling}

As Konz and Benford state~\cite{konz2003}, the angular momentum of an
electromagnetic field is typically on the order of $\hbar$ per photon. In
the case of a circularly polarized beam, $L = \pm\hbar$ per photon, so
$L = P/\omega$ where $P$ is the power, and $\omega$ is the angular
frequency. This clearly shows why, \emph{cetera paribus}, lower frequencies are
more efficient for exerting torque on scatterers.

However, another important
consideration is the fraction of the beam intercepted by the scatterer. The
minimum width of a beam of given frequency is on the order of the wavelength,
so we can consider the minimum cross-sectional area of the beam to be on
the order of $\lambda^2$. Thus, the irradiance $I$ is
\begin{equation}
I \approx P/\lambda^2
\end{equation}
and the power actually incident on the scatterer smaller than the beam is
\begin{equation}
P_\mathrm{inc} \approx Pd^2/\lambda^2
\end{equation}
where $d$ is the scatterer size. The angular momentum incident on the
scatterer will be on the order of
\begin{equation}
L_\mathrm{inc} \approx Pd^2/2\pi c\lambda
\end{equation}
and it can be seen that greater efficiency results from the use of
\emph{shorter} wavelengths.

If the scatterer is larger than the beam, then the scatterer can interact
with the entire incident beam, and the angular momentum incident on the
scatterer will scale as
\begin{equation}
L_\mathrm{inc} \approx P\lambda/2\pi c,
\end{equation}
with greater efficiency at \emph{longer} wavelengths. Accordingly, one
would expect maximally efficient angular momentum transfer when the beam is
focussed to the maximum possible extent, with the wavelength chosen so that
the beam width is on the order of the particle size. This simple argument
is also supported by rigorous electromagnetic calculations~\cite{bishop2003}.

So, while electromagnetic angular momentum generally scales with $1/\omega$,
the transer of this angular momentum to a scatterer does not follow such
simple rules. Rather than lasers being unable to usefully spin objects,
as claimed by Konz and Benford~\cite{konz2003}
optical frequencies are in fact optimal for particles of sizes comparable
to optical wavelengths; particles typically
rotated within laser traps are of this size~\cite{bishop2003}. It can also
be noted that since Konz and Benford~\cite{konz2003} used a beam several
wavelengths wide, they could have obtained greater efficiency by using a
longer wavelength combined with a more strongly focussed beam, unless the
increase in wavelength is accompanied by a sufficient change in the
electromagnetic properties of the material. Of course, the added complication
of a strongly focussed beam might well make this impractical.

\section{Angular momentum of a circularly polarized plane wave}

Konz and Bedford state that a ``circularly polarized incoming wave field
which is infinitely extended and homogeneous does not carry angular
momentum, since the Poynting flux is parallel to the wave vector.''
While a naive calculation of the angular momentum density or flux
starting from $\vec{r}\times\vec{S}$, where $\vec{S}$ is the Poynting
vector gives a result of zero~\cite{khrapko2001},
consideration of the interaction between the
field and an absorbing or anisotropic medium shows that torque is
exerted on the medium~\cite{poynting,beth,feynman,allen2002,yurchenko2002}.
This clearly demonstrates that the wave has non-zero angular
momentum.

While this issue is peripheral to the paper being commented on, recognition
of the fact that a circularly polarized plane wave does carry angular
momentum make the authors' conclusion that the wedge ``absorbs
the negative angular momentum $L_x$ of the reflected wave
field''~\cite{konz2003} unnecessary.

\section{A comment on terminology}

The terminology ``geometric absorption'' of angular momentum
is perhaps an unfortunate choice,
as fundamentally similar processes will act to transfer angular momentum
to a scatterer regardless of the angular momentum content of the incident
field. The authors introduce the awkward concept of absorption of negative
angular momentum from the \emph{scattered} field to account for the angular
momentum transfer to the scatterer in the case of an incident field with
zero angular momentum; this, however, introduces the surprising concept
of absorbing some property of an outward-propagating field.
It would seem to be better to use a term such
as ``geometric transfer'' of angular momentum, so as to eliminate any
conceptual difficulty associated with the ``absorption'' of a quantity
equal to zero.

In addition, to describe negative ``absorption'' of angular momentum as
``radiation'' of angular momentum invites confusion with work involving
pure radiation of angular momentum (with no incident field
present)~\cite{chute1967}.

\end{document}